\documentclass[showpacs,twocolumn,aps,prl,superscriptaddress]{revtex4}

\usepackage{mathrsfs}
\usepackage{amsmath}
\usepackage{amsfonts}
\usepackage{amssymb}
\usepackage{amsthm}
\usepackage{graphicx}

\begin{document}
\title{Anomalous spin Hall and inverse spin Hall effects in magnetic systems}

\author{X. R. Wang}
\email{phxwan@ust.hk}
\affiliation{Physics Department, The Hong Kong University of
Science and Technology, Clear Water Bay, Kowloon, Hong Kong}
\affiliation{HKUST Shenzhen Research Institute, Shenzhen 518057, China}

\begin{abstract}
Anomalous spin-Hall effects (SHE) and anomalous inverse spin-Hall effects (ISHE) are 
predicted when the order parameters are involved in the charge-spin interconversion. 
A spin current whose propagation and polarization are collinear and transverse to an 
applied charge current can be generated when the charge current is parallel to the 
magnetization $\vec M$ in a ferromagnet or to the Neel order $\vec n$ in an 
anti-ferromagnet. When the applied charge current is perpendicular to the order 
parameter, two spin currents can be generated. One is the spin current polarized 
along the order parameter and propagating along the charge current direction. 
The other is the one polarized in the charge current direction and propagating 
along the order parameter direction. Charge currents proportional to the order 
parameter are generated by applied spin currents in the anomalous ISHE. 
When a spin current whose propagation and polarization are collinear is applied, 
the charge current is along the perpendicular (to the spin current) component 
of order parameter, and no current for order parameter along the spin current. 
For applied spin current with propagation and polarization mutually orthogonal to each 
other, a charge current along the spin current propagation direction is generated 
if the order parameter is collinear with the polarization of the spin current. 
A charge current along the polarization of spin current is generated if the 
order parameter is collinear with the propagation direction of the spin current. 
Experimental verifications are also proposed. In terms of applications, one of 
the great advantages of anomalous SHE is that one can control the generated spin or 
charge current by controlling the magnetization or the Neel order in the magnetic 
materials. 
\end{abstract}

\maketitle


Spin current generation, manipulation, and detection are essential themes of 
spintronics. Similar to electric-currents in electronics that perform all kinds 
of tasks, spin currents can carry and process information in spintronics, as 
well as control spin and magnetization direction of memory cells \cite
{sotliu,sot} and magnetic domain wall motion \cite{yanpeng,hubin,xiansi,lqliu}. 
Different from an electric current that is the flow of charges, a spin current 
is the flow of a vector quantity of angular momenta. In the Cartesian 
coordinates, spin currents are tensor of rank two with nine components, 
instead of a vector for electric currents. A spin current can exert a torque 
on spins or on magnetic moments when passing through them, and this torque is 
component sensitive and useful as a control knob for manipulating spin states 
of nano-devices \cite{yanpeng,hubin,xiansi,lqliu}. 

Spin currents can be generated in many systems and in many different ways \cite
{yanpeng,hubin,xiansi,lqliu}, but one very attractive method is through the 
spin Hall effect (SHE) originally proposed by Dyakonov and Perel \cite{SHE1}. 
Recent escalating research activities on SHE and the terminology is largely 
due to the rediscovery of the effect by Hirsch \cite{hirsch} and due to the 
subsequently experimental verifications in semiconductor devices \cite{SHE2,SHE3}. 
This well-known SHE says that a charge current $\vec J$ in a material with 
spin-orbit interaction can generate a spin current $j^s_{ij}$ polarizing in  
direction $\hat j$ and flowing in direction $\hat i$ with $\hat i$, $\hat j$, 
and $\vec J$ mutually perpendicular with each other. Thus, in the Cartesian 
coordinate this spin current is $j^s_{ij}=\theta_0 \hbar/(2e)\epsilon_{ijk} J_k$, 
where $\epsilon_{ijk}$ is the Levi-Civita symbol and the Einstein summation 
convention is applied. $\theta_0$ is a material parameter called spin Hall angle
whose value is an issue of recent debates. Like action-reaction principle in 
nature, the inverse effect of SHE is called inverse spin Hall effect (ISHE) 
\cite{saitoh,kimura,kajiwara} that says a spin current can generate a 
charge current of $J_k=\theta_0'(2e)/(\hbar) \epsilon_{ijk}j^s_{ij}$.
Again the Einstein summation convention is applied and $\theta_0'$ is 
called inverse spin Hall angle, and interestingly, $\theta_0'=\theta_0$. 
This conventional SHE was confirmed capable of reversing spins of spin-orbital 
torque (SOT) memory \cite{sotliu} while the conventional ISHE becomes a  
standard technique for spin current detection \cite{saitoh,kimura,kajiwara}. 
From application point of view, one of the shortcomings of the spin current 
generated from the conventional SHE is that spin-current polarization must be 
perpendicular to the spin-current propagation direction. It is highly desirable 
to generate spin currents whose polarization can be in any possible directions 
and controllable by other external means so that resulted torques are tunable. 
This is the subject of the present study. 

In this letter I predict the existence of new types of charge-spin interconversion 
in magnetic materials. Differ from the existing paradigm \cite{SHE1,hirsch,saitoh,
kimura}, I show, from the general requirement of a genuine physical quantity 
being a tensor, that there are, in principle, three new SHEs and ISHEs in 
magnetic materials when they involve the order parameters of magnetization 
$\Vec M$ in ferromagnets and the Neel order $\vec n$ in anti-ferromagnets.
I term them anomalous SHE and anomalous ISHE because they are linear in the order 
parameter. Specifically, spin currents $j^s_{ii}$ are converted from a charge 
current collinear with the order parameter where $\hat i$ is perpendicular to 
the charge current. When a charge current flows perpendicularly to the order 
parameter, two new spin currents $j^s_{ij}$ are generated, where directions 
$\hat i$ and $\hat j$ are either respectively along electric current and order 
parameter or respectively along the order parameter and electric current. 
In the anomalous ISHE, a charge current along the order parameter is generated 
when a spin current $j^s_{ii}$ with its polarization and propagation collinear 
flow perpendicularly to the order parameter. A charge current along either the 
spin current flow direction or polarization direction is generated by a spin 
current $j^s_{ij}$ when $\hat i$ and $\hat j$ are mutually perpendicular to 
each other, and either $\hat i$ or $\hat j$ is along the order parameter.  

\begin{figure}[htbp]
\centering
\includegraphics[width=0.5\textwidth]{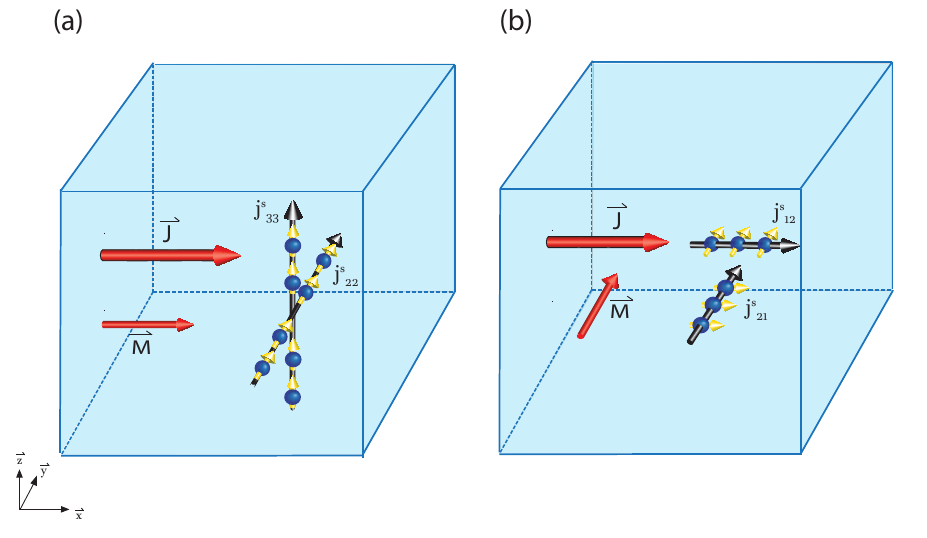}\\
\caption{Schematics of anomalous SHE: (a) The charge current $\vec J$ (the thicker 
red arrowed line) and magnetization $\vec M$ (the thinner red arrowed line) are 
along the $\hat x$-direction. Two anomalous spin current $j^s_{22}$ and $j^s_{33}$ 
(the black arrowed lines) proportional to $M$ and $j$ are generated. 
(b) The charge current $\vec J$ (the thicker red arrowed line) and magnetization 
$\vec M$ (the thinner red arrowed line) are respectively along the $\hat x$- and 
the $\hat y$-directions. Two anomalous spin current $j^s_{12}$ and $j^s_{21}$ (the 
black arrowed lines with polarization decorated by the bead-arrow on the lines) 
proportional to $MJ$ are generated.
}
\label{fig1}
\end{figure}

Consider a piece of ferromagnetic or anti-ferromagnetic metal with an applied 
charge current density $\vec J$. I will take a ferromagnet as an example below.
A spin current $j^s_{ij}$ is generated by $\vec J$. Since spin current is a tensor 
of rank 2 and $\vec J$ a tensor of rank 1, the most general relationship between 
$j^s_{ij}$ and $\vec J$ \cite{yin1,yin2,yin3}, in the linear response region, is  
\begin{equation}
j^s_{ij}=\frac{\hbar}{2e}\theta_{ijk}^\mathrm{SH}J_k,
\label{sh}
\end{equation}
where $\theta_{ijk}^\mathrm{SH}$ is the spin-Hall angle tensor of rank 3 that 
does not depend on current $\vec J$. In three dimension, $i,j,k=1,2,3$ stand for 
respectively the $x$, $y$, and $z$ directions. In the absence of magnetic field, the 
only available non-zero rank tensors (other than the electric current) are order 
parameter $\vec M$ or $\vec n$ for a ferromagnet or an anti-ferromagnet, as well 
as the Levi-Civita symbol of $\epsilon_{ijk}= 1$ for $(i,j,k)=(1,2,3),\ (2,3,1),
\ (3,1,2)$, $\epsilon_{ijk}=-1$ for $(i,j,k)=(1,3,2,),\ (2,1,3,),\ (3,2,1)$, and 
$\epsilon_{ijk}=0$ for any other choices of $(i,jk)$. If the order parameter can 
also participate in the SHE, and if we restrict ourselves to the linear anomalous 
SHE and ISHE in $\Vec M$ or $\vec n$ (in the case of anti-ferromagnet), 
then the most general $\theta_{ijk}^\mathrm{SH}$ in the case of ferromagnet is 
\begin{equation}
\begin{aligned}
\theta_{ijk}^\mathrm{SH}=&\theta_0\epsilon_{ijk} +\theta_1M_l\epsilon_{iln}
\epsilon_{jnk}+\theta_2M_l\epsilon_{ink}\epsilon_{jln},
\label{gen-sh}
\end{aligned}
\end{equation}
$\theta_0$ is the usual spin Hall angle that does not interact with 
$\vec M$, $\theta_\alpha$ ($\alpha=1,2$) are anomalous SHE coefficients.  
It is straight forward to see that the last two terms can be recast as 
\begin{equation}
[(\theta_1+\theta_2)\delta_{ij}\delta_{kl\neq i}+\theta_1\delta_{ik}
\delta_{jl\neq i}+\theta_2\delta_{il}\delta_{jk\neq i}]M_l.
\label{new-sh}  
\end{equation}
In order to see what these anomalous SHEs are, we consider two cases: 
(1) The charge current is along $\vec M$. (2) The 
charge current is perpendicular to $\vec M$. Without losing generality, 
let $\vec J$ and $\vec M$ be along the $\hat x$-axis in case (1). 
Substituting Eqs. \eqref{gen-sh} and \eqref{new-sh} into Eq. \eqref{sh}, 
the general SHE in magnetic materials becomes
\begin{equation}
j^s_{ij}=\frac{\hbar}{2e}[\theta_0\epsilon_{ij1}j
+(\theta_1+\theta_2\delta_{ij\neq 1}) MJ].
\label{new-sh1}
\end{equation}
In case (1), spin current $j^s_{22}$ and $j^s_{33}$ proportional to $MJ$ are 
generated when $\vec J$ and $\vec M$ are collinear and along the $\hat x$-axis. 
Fig. \ref{fig1}(a) is the schematic diagram of the anomalous SHE where the 
thicker and thinner red arrowed lines denote respectively the charge current 
source and magnetization directions. The black arrowed line denote two two 
spin currents whose polarization are indicated by decorated smaller arrows. 
Clearly, the conventional SHE cannot generate such a spin current.

Without losing generality, we let $\vec J$ and $\vec M$ respectively along 
the $\hat x$- and the $\hat y$-directions in case (2). 
From Eqs. \eqref{sh}-\eqref{new-sh}, we have 
\begin{equation}
j^s_{ij}=\frac{\hbar}{2e}(\theta_0\epsilon_{ij1}+\theta_1
\delta_{i1}\delta_{j2}M+\theta_2\delta_{i2}\delta_{j1}M)J.
\label{new-sh2}
\end{equation}
Eq. \eqref{new-sh2} says that spin current $j^s_{12}$ and $j^s_{21}$, proportional to 
the magnetization $M$ and propagating respectively along the $\hat x$- and the $\hat 
y$-directions, are generated. This is different from the conventional SHE that does 
not allow a spin current polarized or propagating along the charge current direction.
The schematic of the new anomalous SHE is shown in Fig. \ref{fig1}(b). 

\begin{figure}[htbp]
\centering
\includegraphics[width=0.5\textwidth]{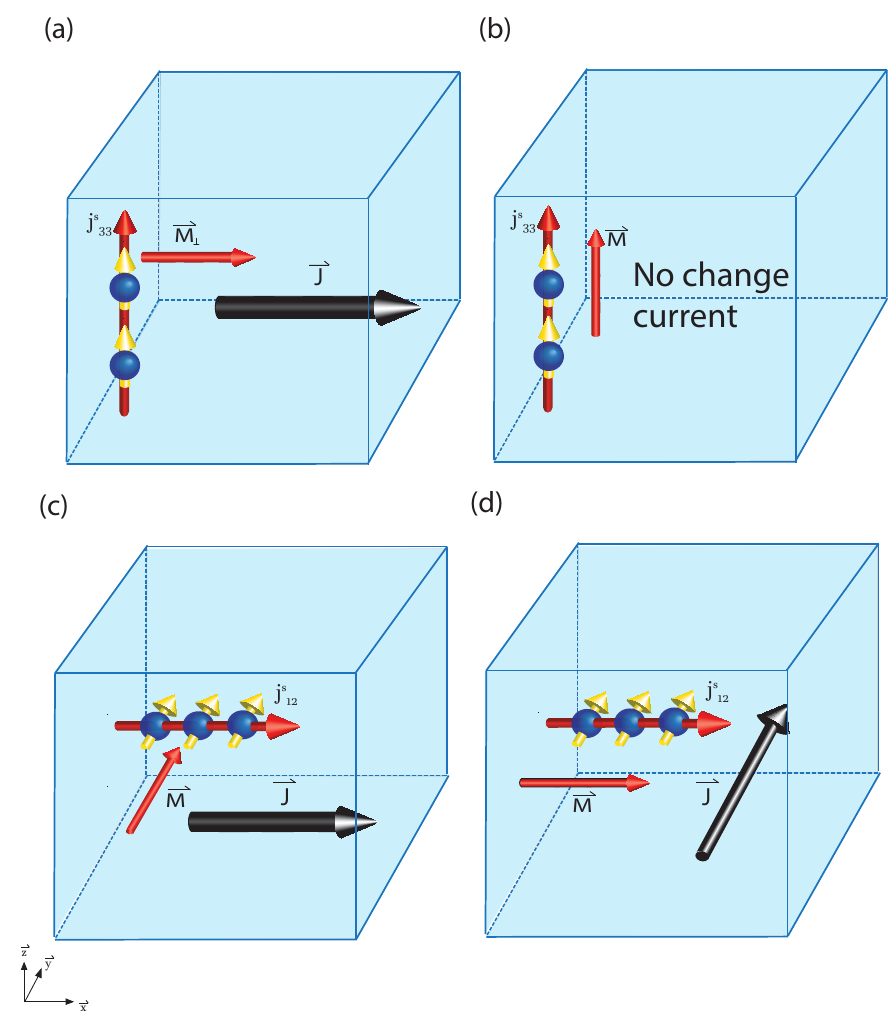}\\
\caption{Schematics of anomalous ISHE: (a) The flow of a spin current $j^s_{33}$ 
(the thinner red arrowed line) generates a charge current along $\vec M_\perp$ 
(the thicker red arrowed line), the projection of $\vec M$ in the $xy$-plane. 
The charge current $\vec J$ (the black arrowed line) is proportional to $M_\perp$. 
(b) No charge current can be generated when $\vec M$ is along the $\hat z$. 
(c-d) Charge currents (the black arrowed lines) proportional to the $y$ (c) and 
$x$ (d) components of the magnetization (the thicker red arrowed lines) are 
generated by the flow of a spin current $j^s_{12}$ (the red thinner lines). 
}
\label{fig2}
\end{figure}

Similarly, an external spin current $j^s_{ij}$ passing through a ferromagnet can 
in principle generate a charge current $\vec J$. By the same arguments, the most 
general expression for $\vec J$ is 
\begin{equation}
J_k=\frac{2e}{\hbar}\theta_{ijk}^\mathrm{ISH}j^s_{ij}
\label{ish}
\end{equation}
where $\theta_{ijk}^\mathrm{ISH}$ is the inverse spin-Hall angle tensor of rank 
3 that does not depend on spin current $j^s_{ij}$. 
For a ferromagnet whose magnetisation can also participant in the generalized 
ISHE, the most general $\theta_{ijk}^\mathrm{ISH}$, up to the linear term 
in $\vec M$, is 
\begin{equation}
\begin{aligned}
\theta_{ijk}^\mathrm{ISH}=&\theta_0^\prime\epsilon_{ijk}
+\theta_1^\prime M_l\epsilon_{iln}\epsilon_{jnk}
+\theta_2^\prime M_l\epsilon_{ink}\epsilon_{jln},
\label{gen-ish}
\end{aligned}
\end{equation}
$\theta_0^\prime$ is the usual inverse spin Hall angle that does not 
interact with $\vec M$, $\theta_\alpha^\prime$ ($\alpha=1,2$) are 
coefficients that characterize the anomalous ISHEs linear in $\vec M$. 
If the applied spin current is $j^s_{33}$, the conventional ISHE says no 
electric current can be generated. However, it is very different here, and 
one has, from Eqs. \eqref{ish} and \eqref{gen-ish},  
\begin{equation}
J_k=\frac{\hbar}{2e}(\theta_1^\prime+\theta_2^\prime)\delta_{ki\neq 3}
M_i j^s_{33}.
\label{new-ish1}
\end{equation}
In this case, the theory predicts that a charge current of $\frac{\hbar}{2e}
(\theta_1^\prime+\theta_2^\prime)j^s_{33}\vec{M_{\perp}}$ 
is generated, where $\vec{M_{\perp}}$ is projection of magnetisation 
vector in the $xy$-plane. The charge current should be zero if the magnetisation 
is parallel to the spin current propagation direction. The schematic of this 
anomalous ISHE is shown is Fig. \ref{fig2}(a).
If the applied spin current is $j^s_{12}$, then charge current from the ISHE 
of Eqs. \eqref{ish} and \eqref{gen-ish} is 
\begin{equation}
J_k=\frac{\hbar}{2e}j^s_{12}[\theta_0^\prime\delta_{k3}+
\theta_1^\prime M_2\delta_{k1}+\theta_2^\prime M_1\delta{k2}].
\label{new-ish2}
\end{equation}
Interestingly, beside of charge current $\frac{\hbar}{2e}j^s_{12}\theta_0
^\prime \hat z$ along the $\hat z$-direction from conventional ISHE, 
there are two new charge currents $\frac{\hbar}{2e}j^s_{12}\theta_1^\prime
M_2\hat x$ and $\frac{\hbar}{2e}j^s_{12}\theta_2^\prime M_1\hat y$ along 
the $\hat x$- and $\hat y$-directions respectively. This two new anomalous 
ISHEs are illustrated in Figs. \ref{fig2}(c) and \ref{fig2}(d). 
No anomalous charge current is generated when $\vec M$ is along the $z$-direction.
Of course, the conventional ISHE can generate a charge current along the $z$ 
direction, and this charge current does not depend on $\vec M$.

The anomalous SHEs and ISHEs described above work also for the interconversion 
of charge-spin in an anti-ferromagnet involving the Neel order parameter $\vec n$. 
One needs simply to replace the magnetization $\vec M$ by $\vec n$. 
In fact, the theory can even be applied to fictitious order parameters such as 
topological insulators or 2D materials whose surface direction involve in many 
physical phenomenon. Indeed, there are already many evidences for the existence 
of charge-spin conversion beyond current paradigm of SHE. For example, the 
existence of anomalous SHE may have already observed in ferromagnet 
\cite{Jungwirth}, in anti-ferromagnets \cite{Otani,yang,song} and in 
two-dimensional (2D) Weyl semimetals (a case of anomalous SHE involved 
fictitious order parameters of 2D materials) \cite{ralph1,ralph2}. 
Of course, it shall be interesting to test the predictions made here quantatively. 

\begin{figure}[htbp]
\centering
\includegraphics[width=0.45\textwidth]{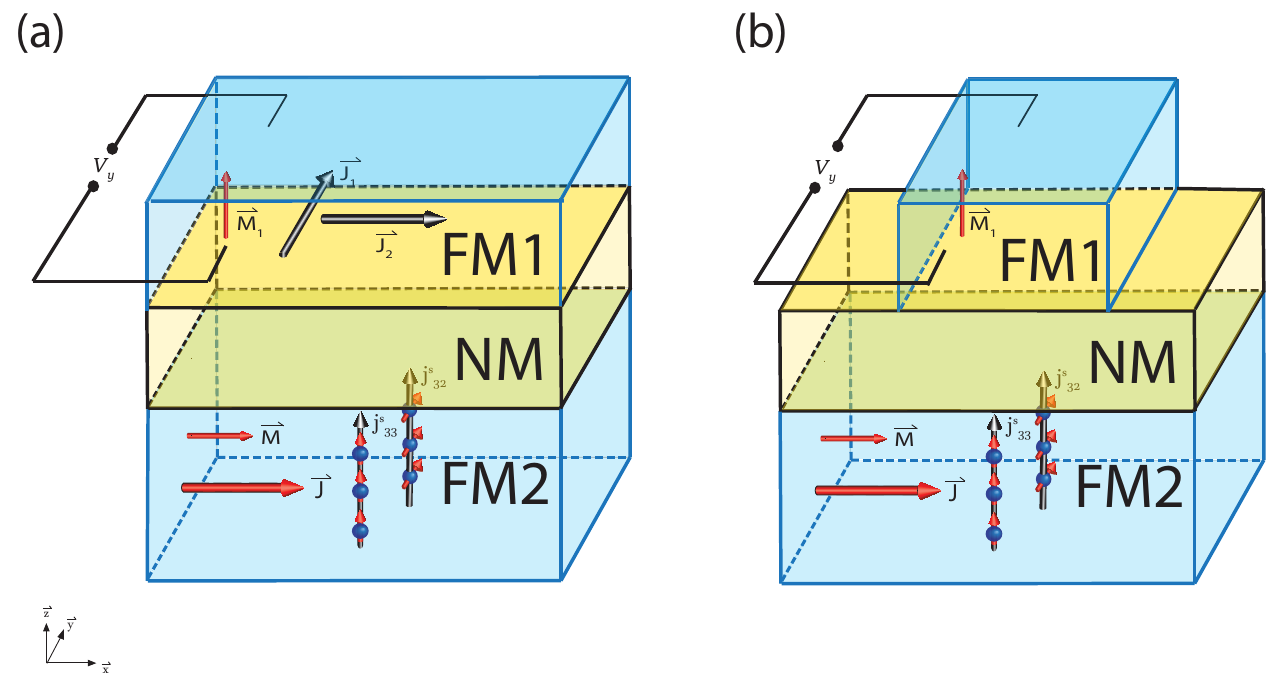}\\
\caption{(Color online) Illustration of an experimental set-up of FM1/NM/FM2 
multilayer system lying in the $xy$-plane. FM2 is a ferromagnetic metal layer 
of magnetization $\vec M$ (the thinner red arrowed line) for spin current generation. 
Applied charge current $\vec J$ (the thicker red arrowed line) is collinear with 
$\vec M$ and along the $x$-direction. NM is non-magnetic-spacing layer that can be 
either a metallic or a thin insulating film. FM1 is another ferromagnetic metal for 
spin current detection whose magnetization $\vec M_1$ is along the $z$-direction. 
Spin currents $j^s_{33}$ and $j^s_{32}$ (the black arrowed lines decorated with 
smaller arrows) are generated by the anomalous ($j^s_{33}$) and conventional SHE. 
The two spin currents passing thought the spacing layer and enter into the top 
ferromagnetic metal layer FM1. $\vec J_1$ and $\vec J_2$ are two charge currents 
generated from the anomalous ISHE.
A voltage drop $V_y$ across the $y$ direction in FM1 is generated. (a) A set-up 
for NM being a thin insulating spacing. (b) For NM being a metal, FM1 is replaced 
by a Hall bar along the $y$ direction. 
}
\label{fig3}
\end{figure}

Here we would like to propose one possible experimental set-up for experimental  
verification of predictions described above. Figure \ref{fig3} is the 
illustration of the set-up. A FM1/NM/FM2 multilayer system lays in the $xy$-plane, 
and a charge current $\vec J$ is applied to the bottom ferromagnetic layer FM2  
whose magnetization magnetization $\vec M$ is collinear with $\vec J$.
NM is non-magnetic-spacing layer that can be metal or thin insulating film.
The charge current direction is chosen as the $x$-direction.
The charge current generate two spin currents $j^s_{33}=(\theta_1+\theta_2)
\frac{\hbar}{2e}MJ$ and $j^s_{32}=\theta_0\frac{\hbar}{2e}J$ (the 
black arrowed lines) propagating perpendicularly to the layers.
One of them $j^s_{32}$ is the charge current from the conventional SHE.
The other $j^s_{33}$ is from the anomalous SHE. The two spin currents pass 
thought the spacing layer and enter into the top ferromagnetic metal layer FM1. 
If magnetization $\vec M_1$ in FM1 is along the $z$-direction, $j^s_{33}$ 
generates no charge current from anomalous ISHE illustrated in Fig. 
\ref{fig2}b, as well as no current from the conventional ISHE 
because the propagation and polarisation of the current are collinear. 
Spin current $j^s_{32}$ can generate two charge current in FM2. 
One of them $\vec J_1=\theta_0^\prime\frac{2e}{\hbar}j^s_{32}$ is from 
the conventional ISHE, and is flow in the $x$-direction. 
The other $\vec J_2=\theta_2^\prime\frac{2e}{\hbar}M_1j^s_{32}$ proportional
to $M_1$ is from the anomalous ISHE that is explained in Fig. \ref{fig2}d. 
For an open circuit in the $y$-direction on the top layer (FM1), it shall 
generate a voltage drop cross the $y$-direction as denoted by $V_y$ in Fig. 
\ref{fig3}. $V_y$ shall change sign when $\vec M_1$ reverses its direction. 
Both the existence of a non-zero $V_y$ and its sign change are the features of 
the anomalous ISHE. Ideally, one would like to eliminate closed circuit in the 
$x$-direction on the top layer in order to eliminate possible contribution of 
usual anomalous Hall contribution to $V_y$. This can be achieved either by using 
a thin insulating spacing or by using a Hall bar as denoted by Fig. \ref{fig3}b. 
Obviously, the charge current along the $y$-direction is a uniquely feature of 
the anomalous ISHE predicted in this theory. Because all existing effects do 
not result in a $\vec M_2$ dependent $V_y$. 

According to the anomalous SHE and ISHE described above, a charge current along 
the x-directions in a ferromagnetic or anti-ferromagnetic metal, whose order 
parameter is collinear with the current, can generate $j^s_{yy}$ and $j^s_{zz}$. 
Since the $y$ and $z$ direction is open and no sustainable spin current exist, 
a spin accumulation of $<s_z>$ and $<s_y>$ will occur on the two surface 
respectively in the $z$ and $y$ directions. Thus this direction-dependent 
spin accumulation is another fingerprints of the present theory. 
Of course, this spin accumulation is hidden in the main order parameter 
pointing to the $x$-direction, and how to observe it may be an issue. 
It may be easier to observe the spin accumulation in an anti-ferromagnet 
because of zero net magnetization everywhere in the absence of a current.  

Clearly, all three anomalous SHEs and ISHEs involve the coupling between order 
parameters such as the magnetization of ferromagnets or the Neel order of 
anti-ferromagnets. In terms of applications, this nice property allow one to 
manipulate the generated spin or charge currents by controlling the order parameter. 
So far, the predictions are based on the general principle of tensor 
transformation, instead of deriving these terms from a microscopic Hamiltonian. 
In some sense, similar to energy spectrum analysis in group theory, our theory 
does not provide the actual possible strengths of the anomalous SHEs and 
ISHEs, $\theta_1$, $\theta_1^\prime$, $\theta_2$, and $\theta_2^\prime$. 
To find these strengths, one needs to compute all possible spin (charge) 
current conversion from a given microscopic model when an external charge 
(spin) current is applied. Such a microscopic theory is surely important 
and necessary although it is foreseeable difficult because of necessity 
of including electron-electron interactions.   

In conclusion, I predict the existence of anomalous SHEs and ISHEs in 
magnetic materials when the charge-spin interconversion involve the 
order parameters such as the magnetization in ferromagnetic materials
and the Neel order in anti-ferromagnetic materials. In particular, 
spin current $j^s_{ii}$ can be generated by a charge current collinear 
with the order parameter and propagating perpendicularly to $\hat i$. 
Inversely, a charge current can be generated by a spin current of $j^s_
{ii}$ along the projection of order parameter perpendicular to $\hat i$.
Two anomalous spin currents proportional to the magnitude of order parameter 
can be generated by an applied charge current when it is perpendicular to 
the order parameter. One of them propagates along the order parameter and
is polarized along the charge current direction. The other propagates along 
the charge current direction and is polarized along the order parameter. 
Its inverse effect is the generation of two charge currents by a spin current 
of $j^s_{ij\neq i}$. One current proportional to the $i$'th component of the 
order parameter flow along the $\hat j$-direction, and the other proportional to 
the $j$'th component of the order parameter flow along the $\hat i$-direction.

This work was supported by the National Natural Science Foundation of 
China (Grant No. 11774296 and 11974296) as well as Hong Kong RGC Grants No. 
16301518 and No. 16301619. I thank Xuchong Hu for preparing the figures.


\begin{thebibliography}{99}

\bibitem{sotliu} Luqiao Liu, Chi-Feng Pai, Y. Li, H. W. Tseng, D. C. Ralph, 
R. A. Buhrman, ``Spin-Torque Switching with the Giant Spin Hall Effect of 
Tantalum", \textit{Science} \textbf{336} (6081), 555-558 (2012).

\bibitem{sot} Y. Zhang, H. Y. Yuan, X. S. Wang, and X. R. Wang, ``Breaking 
the current density threshold in spin-orbit-torque magnetic random access 
memory", \textit{Phys. Rev. B} \textbf{97}, 144416 (2018).

\bibitem{yanpeng} P. Yan, X. S. Wang, and X. R. Wang, ``All-magnonic 
spin-transfer torque and domain wall propagation", \textit{Phys. 
Rev. Lett.} \textbf{107}, 177207 (2011).

\bibitem{xiansi} X.S. Wang, P. Yan, Y. H. Shen, G.E.W. Bauer, and
X.R. Wang, ``Domain wall propagation through spin wave emission", 
\textit{Phys. Rev. Lett.} \textbf{109}, 167209 (2012).

\bibitem{hubin} B. Hu and X. R. Wang, ``Instability of Walker Propagating Domain 
Wall in Magnetic Nanowires", \textit{Phys. Rev. Lett.} \textbf{111}, 027205 (2013).

\bibitem{lqliu} Jiahao Han, Pengxiang Zhang, Justin T. Hou, Saima A. Siddiqui, 
and Luqiao Liu, ``Mutual control of coherent spin waves and magnetic domain walls 
in a magnonic device", \textit{Science} \textbf{366} (6469), 1121-1125 (2019). 

\bibitem{MacDonald}D. K. C. MacDonald and K. Sarginson, ``Galvanomagnetic 
effects in conductors", \textit{Repts. Progr. in Phys.} \textbf{15}, 249 (1952).

\bibitem{conwell}E. M. Conwell,``Galvanomagnetic effects in semiconductors
at high electric fields, \textit{Phys. Rev.} \textbf{123}, 454 (1961).

\bibitem{ganichev}S. D. Ganichev, E. L. Ivchenko, V. V. Bel'kov, 
S. A. Tarasenko, M. Sollinger, D. Weiss, W. Wegscheider and W. Prettl, 
``Spin-galvanic effect", \textit{Nature} \textbf{417}, 153 (2002).

\bibitem{QHE}\textit{The Quantum Hall Effect}, edited by R. E. Prange
and S. M. Girvin (Springer-Verlag, New York, 1990).

\bibitem{pugh}Emerson M. Pugh and Norman Rostoker, ``Hall effect in
ferromagnetic materials", \textit{Rev. Mod. Phys.} \textbf{25}, 151 (1953).

\bibitem{SHE1}M. I. Dyakonov and V. I. Perel, ``Possibility of orientating 
electron spins with current". Sov. Phys. JETP Lett. \textbf{13}, 467 (1971). 

\bibitem{hirsch}J. E. Hirsch, ``Spin Hall effect",
\textit{Phys. Rev. Lett.} \textbf{83}, 1834 (1999).

\bibitem{SHE2}Y. Kato; R. C. Myers; A. C. Gossard; D. D. Awschalom, 
``Observation of the Spin Hall Effect in Semiconductors". \textit{Science}
\textbf{306}(5703), 1910-1913 (2004). 

\bibitem{SHE3}J. Wunderlich; B. Kaestner; J. Sinova; T. Jungwirth, 
``Experimental Observation of the Spin-Hall Effect in a Two-Dimensional 
Spin-Orbit Coupled Semiconductor System", \textit{Phys. Rev. Lett.} 
\textbf{94}(4), 047204 (2005). 

\bibitem {saitoh}E. Saitoh, M. Ueda, H. Miyajima and G. Tatara, ``Conversion 
of spin current into charge current at room temperature: inverse spin-Hall 
effect", \textit{Appl. Phys. Lett.} \textbf{88}, 182509 (2006).

\bibitem{kimura}T. Kimura, Y. Otani, T. Sato, S. Takahashi, and S. Maekawa,
``Room-temperature reversible spin Hall effect",
\textit{Phys. Rev. Lett.} \textbf{98}, 156601 (2007).

\bibitem{kajiwara}Y. Kajiwara, K. Harii, S. Takahashi, J. Ohe, K. Uchida,
M. Mizuguchi, H. Umezawa, H. Kawai, K. Ando, K. Takanashi, S. Maekawa and 
E. Saitoh, ``Transmission of electrical signals by spin-wave interconversion 
in a magnetic insulator", \textit{Nature} \textbf{464}, {262}-{267} (2010).

\bibitem{yin1}Y. Zhang, H. W. Zhang, and X. R. Wang, ``Extraordinary 
galvanomagnetic effects in polycrystalline magnetic films", 
\textit{Europhys. Lett.} \textbf{113}, 47003 (2016).

\bibitem{yin2} Y. Zhang, X. S. Wang, H. Y. Yuan, S. S. Kang, H. W. Zhang, 
and X. R. Wang, ``Dynamic magnetic susceptibility and electrical detection of 
ferromagnetic resonance", \textit{J. Phys.: Condens.  Matter} \textbf{29}, 095806 (2017).

\bibitem{yin3}Y. Zhang, Q. Liu, B. F. Miao, H. F. Ding, and X. R. Wang, 
``Anatomy of electrical signals and dc-voltage line shape in spin-torque 
ferromagnetic resonance", \textit{Phys. Rev. B.} \textbf{99}, 064424 (2019).


\bibitem{Jungwirth}H. Kurebayashi, J. Sinova, D. Fang, A. C. Irvine, T. D. Skinner, 
J. Wunderlich, V. Novak, R. P. Campion, B. L. Gallagher, E. K. Vehstedt, L. P. Zarbo, 
K. Vyborny, A. J. Ferguson, and T. Jungwirth, ``An antidamping spin-orbit torque 
originating from the Berry curvature", \textit{Nat. Nanotech.} \textit{9}, 211–217 (2014).

\bibitem{Otani}M. Kimata, H. Chen, K. Kondou, S. Sugimoto, P. K. Muduli, 
M. Ikhlas, Y. Omori, T. Tomita, A. H. MacDonald, S. Nakatsuji, and Y. Otani, 
``Magnetic and magnetic inverse spin Hall effects in a non-collinear 
antiferromagnet", \textit{Nature} \textit{565}, 627 (2019).

\bibitem{yang}Y. Liu, Y. Liu, M. Chen, S. Srivastava, P. He, K. L. Teo, T. 
Phung, S-H. Yang, and H. Yang "Current-induced out-of-plane spin accumulation 
on the (001) surface of the IrMn3 antiferromagnet", \textit{Phys. Rev. Appl.} 
\textit{12}, 064046 (2019).

\bibitem{song}X. Chen, X. Zhou, R. Cheng, C. Song, J. Zhang, Y. Wu, Y. Ba, 
H. Li, Y. Sun, Y. You, Y. Zhao, and F. Pan, ``Electric field control of Néel spin-orbit 
torque in an antiferromagnet", \textit{Nat. Mater.} \textit{18}, 931 (2019).

\bibitem{ralph1}D. MacNeill, G. M. Stiehl, M. H. D. Guimaraes, R. A. Buhrman, 
J. Park and D. C. Ralph, ''Control of spin-orbit torques through crystal symmetry 
in WTe2/ferromagnet bilayers", \textit{Nat. Phys.} \textit{13}, 300 (2017).

\bibitem{ralph2}M. H. Guimaraes, G. M. Stiehl, D. MacNeill, N. D. Reynolds, 
and D. C. Ralph, ``Spin-orbit torques in NbSe2/permalloy bilayers", 
\textit{Nano Lett.} \textit{18}, 1311–1316 (2018).


\end{thebibliography}
\end{document}